\documentclass[12pt]{article}
\usepackage{graphicx}
\usepackage{latexsym}
\usepackage[left=.75in,top=1in,right=.75in,bottom=1in]{geometry}
\usepackage{amsmath}

    \def\independenT#1#2{\mathrel{\setbox0\hbox{$#1#2$}%
    \copy0\kern-\wd0\mkern4mu\box0}} 

\newenvironment{packed_item}{
\begin{itemize}
 \setlength{\itemsep}{0pt}
  \setlength{\parskip}{0pt}
  \setlength{\parsep}{0pt}
}{\end{itemize}}

\usepackage{amssymb}

\usepackage{multirow}
\usepackage[numbers, sort&compress]{natbib}
\usepackage[section]{placeins}

\usepackage[font={footnotesize,it}]{caption}

\usepackage{url}

\usepackage{pdflscape}
\usepackage{arydshln}

\usepackage[dvipsnames]{xcolor}

\usepackage[normalem]{ulem} 
\usepackage[makeroom]{cancel}

\title{Brief Article}
\author{The Author}
\date{November 10, 2021}                                           

\begin{document}

\begin{centering}
\Large{Sustainable East Africa Research in Community Health \\ (SEARCH) Collaboration

\vspace{2em}
STRATEGIC ANTIRETROVIRAL THERAPY AND HIV TESTING FOR YOUTH IN RURAL AFRICA (SEARCH-Youth)

\vspace{3em}

{\textbf{ Statistical Analysis Plan for 
Primary and Selected Secondary Health Endpoints 
of the SEARCH-Youth Study
}}
\vspace{4em}

 Laura B. Balzer, PhD$^1$ \\
 Theodore Ruel, MD$^2$ \\
Diane V. Havlir, MD$^2$ \\
for the SEARCH-Youth Study Team
\vspace{4em}

April 14, 2022 \\
Version 1.0}

\end{centering}
\vfill
$^1$School of Public Health \& Health Sciences, University of Massachusetts Amherst, Amherst, USA $^2$Division of HIV, ID, \& Global Medicine, Department of Medicine, UCSF, San Francisco, USA

\newpage

\tableofcontents 


\section{Brief study overview}

SEARCH-Youth is a cluster randomized trial designed to evaluate the effect of a combination intervention on HIV viral suppression among adolescents and young adults with HIV in rural Uganda and Kenya (Clinicaltrials.gov: NCT03848728). The patient-centered intervention was developed using the PRECEDE model of behavioral change and was designed to address the structural barriers and unique needs of the target population. Details of the trial design and procedures can be found in the Study Protocol. This document provides the statistical analysis plan to evaluate the intervention effect on the primary endpoint of viral suppression and selected secondary endpoints, including mortality, care engagement, and patient satisfaction. Analysis plans for qualitative outcomes, cost-effectiveness, and sub-studies are available elsewhere.   A history of changes to this analysis plan is given in the Appendix.

\subsection{Study design} 
SEARCH-Youth is an implementation science study, designed to rigorously evaluate the effects of a combination intervention, including life-stage assessments, structured choice clinic access, and rapid viral load feedback, on care outcomes among adolescents and young adults (15-24 years) with HIV. Since the personalized intervention is delivered through changes in clinical practice, the unit of randomization is government sponsored health clinics in rural western Kenya and southwestern Uganda. Specifically, a total of 28 clinics (14 per country) were randomized to the SEARCH-Youth intervention or the optimized standard of care: differentiated ART care, standard clinic hours, routine viral load monitoring, and access to 2nd and 3rd line ART. To improve statistical power, randomization occurred within strata defined by country, clinic size, and prior participation in the  SEARCH Study (ClinicalTrials.gov: NCT01864603).
Formal power calculations are given in the Appendix.  

\subsection{Description of population and context} 

We will provide a CONSORT diagram to detail participant flow from recruitment through inclusion in the primary analysis. We will also describe the study population with summary measures of baseline participant characteristics, including
\begin{packed_item}
\item Age
\item Sex
\item Country
\item Education
\item Employment status
\item Marital status
\item Number of children 
\item Alcohol consumption
\item Mobility 
\item ART regimen at enrollment
\item HIV viral suppression, defined as HIV RNA $<$400 cps/mL
\item Care status at enrollment
    \begin{packed_item}
    \item \textbf{Recently engaged:} started ART within the prior 6 months or at enrollment
    \item \textbf{Engaged:} started ART more than 6 months ago, with a clinic visit in the prior 6 months
    \item \textbf{Re-engaging:} started ART more than 6 months ago, without a clinic visit in the prior 6 months
    \end{packed_item}
\end{packed_item}

We will provide these summary measures overall, by arm, by country (Kenya and Uganda), and within country by arm. We will also provide these summary measures at the clinic-level. We will additionally provide descriptions of secular events, non-study activities that could impact study outcomes, and measurement coverage of key endpoints. 

\section{Primary endpoint: Virologic suppression at 2 years}

The primary endpoint is HIV viral suppression (HIV RNA $<$400 cps/mL) at 2 years of follow-up.  Ideally, a HIV RNA measure (hereafter called ``viral load'') would be obtained on every participant on the exact date marking 2 years of follow-up (hereafter called the ``2-year mark''). However, such an approach is logistically impossible and not responsive to study context, specifically the increased mobility expected given the participant’s life-stage and the ongoing COVID-19 pandemic. Therefore, for each participant, the endpoint window will begin 90 days prior to their 2-year mark and end at the nearest of 180 days after their 2-year mark or the date of database closure (March 1, 2022 for the primary endpoint).

Through rigorous study procedures, including clinic visits and home-based tracking, we will attempt to obtain a viral load on all participants during their endpoint window, regardless of their care status. (The exception will be persons who have formally been withdrawn from the study.) During endpoint ascertainment, we will also obtain each participant’s vital status and current residency. We will determine whether each participant lives within the study region, defined as $<$ 3 hours traveling distance within the study country, or has \textbf{outmigrated}, defined as moving outside the study region. We will also review clinical records for documentation of formal transfers of care.

\textbf{Inclusion/exclusion criteria:} All analyses will exclude persons who have been withdrawn from the study. In the primary analysis, we will also exclude participants who have outmigrated or have formally transferred care. The excluded individuals do not represent the trial’s \textbf{target population}: adolescents and young adults with HIV who are living in the catchment area of the study clinics and, thus, able to benefit from the study intervention. Additionally, viral loads obtained on participants who outmigrated are likely to be enriched for persons engaged in care (e.g., they returned to the study clinic during the endpoint window or they transferred care and had a routine viral load available at their new clinic). As described below, secondary analyses will include participants who outmigrated as well as those who transferred. The primary analysis will also restrict to participants enrolled prior to December 1, 2019; secondary analyses will include all participants, pending data availability. 

\textbf{Endpoint classification:} We will classify participants as virally suppressed if their endpoint HIV RNA is $<$ 400 cps/mL. In the primary analysis, we will classify participants as \textbf{failures} if they are unsuppressed (HIV RNA $\geq$400 cps/mL), if they have died prior to their endpoint viral load, or if the study was unable to obtain a viral load in their endpoint window. The reasons for incomplete viral load capture include participant’s declining measurement, participant’s not being found during tracking and without a routine viral load available, or that the sample was obtained after database closure.

\textbf{Clinic-level endpoint:} In the primary analysis, the endpoint (viral suppression or failure) is not subject to missingness. Therefore, in each clinic separately, we will calculate the proportion of participants meeting the inclusion criteria who are suppressing viral replication. In secondary analyses, we will also include participants who outmigrated and participants who formally transferred care. In sensitivity analyses, we will (i) exclude participants without an endpoint viral load measured, (ii) adjust for missingness (detailed below), (iii) treat transfers as successes, and (iv) censor at death.

In analyses subject to missingness or censoring, we will implement targeted minimum loss-based estimation (TMLE) in each clinic separately \citep{MarkBook, Balzer2021twostage}. TMLE allows flexible adjustment for the ways in which participants with measured endpoint viral loads could differ from participants without endpoint viral loads. Our adjustment set will consist of age, sex, and baseline viral suppression status. Our adjustment set will also include a post-baseline indicator of outmigration. To minimize the risk of model misspecification bias, we will use Super Learner, an ensemble machine learning method, to combine predictions from generalized linear models, generalized additive models, and the mean \citep{SuperLearner}. 

For each method, we will use influence curve-based inference to obtain 95\% confidence intervals on the point estimate of the clinic-specific proportion with viral suppression.

\textbf{Effect estimation:} Viral suppression at two-years will be compared between randomized arms using \emph{Two-Stage TMLE}, an approach that appropriately accounts for the dependence of outcomes within clusters \citep{Balzer2021twostage}. In the first stage, viral suppression will be estimated for each clinic, as described above. In the second stage, we will use a cluster-level TMLE to evaluate intervention effectiveness by comparing the clinic-specific endpoints between randomized arms, adaptively selecting the clinic-level adjustment variables that optimize precision, while tightly preserving Type-I error control \citep{Balzer2016DataAdapt,Balzer2021twostage, Benitez2021}. Specifically, we will use leave-one-out cross-validation to select from the following baseline covariates the combination that maximizes empirical efficiency: the clinic-specific number of adolescents and young adults in HIV care at baseline, the clinic-specific proportion with viral suppression among those engaged at baseline, or nothing (unadjusted). In secondary analyses, we will implement an unadjusted effect estimator as the contrast in the average proportion with viral suppression between arms.

The primary effect will be on the relative scale for the N=28 clinics in the trial (i.e., the sample risk ratio) \citep{Neyman1923,Rubin1990,Imbens2004, Imai2008, Balzer2016SATE}: 
\begin{align*}
\frac{\psi(1)}{\psi(0)} = \frac{1/N \sum_{i=1}^N \alpha_i Y_i(1) }{1/N \sum_{i=1}^N \alpha_i Y_i(0)}
\end{align*}
where $\alpha_i$ are weights for the clinic $i$; $Y_i(1)$ is the counterfactual proportion with viral suppression at 2 years if possibly contrary-to-fact clinic $i$ were in the intervention arm ($Z=1$), and $Y_i(0)$ is the counterfactual proportion with viral suppression at 2 years if possibly contrary-to-fact clinic $i$ were in the standard-of-care arm ($Z=0$). In secondary analyses, we will examine the effect on the absolute scale:  $\psi(1)-\psi(0)$. Primary analyses will weight the $N$ clinics equally: $\alpha_i=1$ for all $i$. In sensitivity analyses (described below), we will weight according to the number of participants included in the analysis \citep{Benitez2021}.  

\textbf{Statistical inference:} In all analyses, we will use the influence curve for standard error estimation \citep{Balzer2016SATE, Balzer2021twostage}. All analyses will use the Student's $t$-distribution with 28-2=26 degrees of freedom as finite sample approximation to the standard normal distribution \citep{HayesMoulton2009}. 

We will test the \emph{null hypothesis} that the SEARCH-Youth intervention did not improve viral suppression, as compared to the optimized standard-of-care, with a one-sided test at the 5\% significance level. We will report point estimates and two-sided 95\% confidence intervals for the arm-specific endpoints as well as the relative and absolute effects.

\subsection{Sensitivity analyses} In addition to the previously described sensitivity analyses, we will evaluate the intervention effect using a \emph{Single-Stage TMLE} that pools across clinics to estimate and compare the average of individual-level outcomes between arms \citep{Balzer2018Hierarchical, Balzer2021twostage, Benitez2021}. This approach naturally estimates an individual-level effect (e.g., the by-arm contrast in the overall proportion with viral suppression) and thus weights participants, not clinics, equally. As detailed in \cite{Benitez2021}, however, this effect can equivalently be estimated with Two-Stage TMLE when the appropriate weights are applied. The single-stage approach is preferable when the cluster size is small or the outcome is rare. (The two-stage approach is preferable when the outcome is subject to differential measurement or censoring.)

In sensitivity analyses of the primary endpoint evaluated with the Single-Stage TMLE, we will adaptively select from the following prespecified set the combination of baseline variables that maximizes empirical efficiency: age, sex, viral suppression status, or nothing (unadjusted). In sensitivity analyses of secondary endpoints adjusting for potentially differential measurement of viral loads, we will include these variables as well as the post-baseline variable of outmigration in the adjustment set in a longitudinal TMLE \citep{Petersen2014ltmle, ltmlePackageb}. Throughout statistical inference will account for clustering through influence curve-based estimation, described above. 

As an additional sensitivity analysis, we may define viral suppression as HIV RNA$<$50 cps/mL. 

Finally, given existing programs to support adherence and retention among adolescents and young adults with HIV in study clinics, we may conduct sensitivity analyses to exclude these clinics. We note both intervention and control participants had full access to these programs; the intervention was added to, not in place of, the standard-of-care.

\subsection{Subgroups}
To understand effect heterogeneity, we will repeat the above analyses within the following pre-specified baseline subgroups: country, sex, age group (15-19 years; 20-24 years), baseline care status, baseline ART regimen, and viral suppression status. We anticipate that some of these subgroups will have limited participants per clinic, precluding use of Two-Stage TMLE. Therefore, we will conduct all subgroup analyses using the Single-Stage TMLE, described above. Again, all analyses will account for clustering.

\subsection{Participant flow and endpoint ascertainment}

Recall that the primary analysis excludes participants who outmigrated or transferred care. The remaining participants are then classified as virally suppressed (HIV RNA$<$400 cps/mL) or otherwise (a.k.a., “failures”). To further understand the measurement and endpoints of participants, we will report the number and proportion of participants in the following mutually exclusive and exhaustive categories: 
\begin{packed_item}
\item Outmigrated
\item Transferred care within the study region 
\item Measured viral load (among those not transferred and remaining in the study region) 
\item Died (among those not transferred and remaining in the study region)
\item Otherwise have a missing endpoint viral load 
\end{packed_item}

These metrics will be reported overall and by arm. 

Additionally, for both the primary analytic population (excluding outmigrations and transfers) and secondary analytic population (including outmigrations and transfers), we will provide the arm-specific number and proportion of participants who are suppressed, unsuppressed, dead, or missing at two-years. These raw proportions differ from the primary analysis, which contrasts the average of the clinic-level proportions by arm and adjusts to increase precision.

\subsection{Predictor analyses}

In addition to the previously described subgroup analyses, we will conduct the following analyses to elucidate outcome predictors. As outcomes, we will consider being classified as a ``failure'' (i.e., viral non-suppression, dead, or otherwise missing); having viral non-suppression or dying (adjusting for missingness), and having viral suppression (adjusting for missingness and censoring at death). Sensitivity analyses will exclude participants with missing endpoint viral loads. 

Within each arm separately, we will estimate unadjusted associations and adjusted associations with these outcomes. Baseline predictors will include occupation, marital status, being a parent, drinking alcohol, and mobility. Adjusted analysis will additionally control for country, sex, age, and baseline care group. Adjusted analyses will treat each baseline predictor, in turn, as the exposure variable and the remainder as the adjustment set. All analyses will be implemented with the Single-Stage TMLE and yield associations on the relative scale (statistical analogs of the causal risk ratio).

\section{Impact of the roll-out of dolutegravir} 
During the study follow-up, dolutegravir (DTG) was rolled out across both Kenya and Uganda. According to Ministry of Health guidelines, participants with a recent (within 6 month) viral load $<$1000 cps/mL were eligible to switch to DTG; however, concerns of neural tube defects may have led to delays in switching for women of reproductive age, who represent the majority of our study population. Additionally, DTG is known to be more robust at achieving viral suppression than prior ART regimens. Therefore, we plan to conduct the following analyses to understand the impact of the DTG rollout on study results.  

Overall, by study arm and by sex, we will report the number and proportion of participants who were on DTG at baseline as well as the number and proportion who switched to DTG during study follow-up. Then using the Kaplan-Meier method, we will estimate the longitudinal probabilities of switching to DTG, with censoring at death, endpoint viral load ascertainment, and database closure. Using the two-stage approach and the same approach to covariate adjustment, we will estimate the intervention effect on the cumulative probability of switching by 6, 12, 18, and 24 months. We formally test the corresponding null hypotheses that the probability of switching was equal between arms (relative risk=1) using a two-sided test at the 5\% significance level. Additionally, we will create Kaplan-Meier plots for time-to-switch, by arm, by sex, and by clinic.

Second, we will stratify on switch status and evaluate the intervention effect on viral suppression at 2 years. Among participants who switched to DTG during follow-up, we will test the null hypothesis that the intervention did not improve the probability of viral suppression using a one-sided test at the 5\% significance level. Likewise, among participants who did not switch to DTG during follow-up, we will test the null hypothesis that the intervention did not improve the probability of viral suppression using a one-sided test at the 5\% significance level. 

Since switching is a post-baseline variable and potentially impacted by the intervention, we are losing (i.e., blocking) some of the intervention effect by stratifying on it. In other words, formal mediation analyses are warranted. These analyses would allow us to examine the controlled and natural direct effect of the SEARCH-Youth intervention on viral suppression, accounting for switching to DTG as a mediator \citep{Petersen2006,Rudolph2018}. However, substantial methodological work remains to modify these methods for cluster randomized trials. Additionally, we expect that in the future, most, if not all, participants will be on DTG. Therefore, we will instead formally examine the intervention effect on the joint probability of switching to DTG and attaining viral suppression at 2 years. We will again test the corresponding null hypothesis with a one-sided test at the 5\% significance level.

All analyses will be conducted for the primary analytic population (excluding outmigrations and transfers) and secondary analytic population (including outmigrations and transfers). All analyses will be subset on participants who are not on DTG at baseline.

\subsection{Other potential mediators}

Other potential effect mediators include 
\begin{packed_item}
\item switches to second-line therapy
\item engagement in care, defined as having any clinic contact in the 6 months preceding the endpoint window
\item postpartum status, defined as in two ways: (i) reporting a birth within 6 months of the endpoint viral load measurement, and (ii) reporting a birth within 12 months of the endpoint viral load measurement  
\end{packed_item}

To understand the impact of these post-baseline events on viral suppression and other secondary endpoints (described below), we may conduct analogous analyses as for the switch to DTG.

\section{Secondary health endpoints}

We will evaluate the SEARCH-Youth intervention effect on the following secondary endpoints. For each secondary endpoint (unless noted), a Single-Stage TMLE will be used to evaluate the intervention effect. We will use the same approach to define the primary and secondary analytic populations as for the primary endpoint (viral suppression at 2 years). We will also use the same approach to covariate adjustment, consider the same subgroups, and may implement analogous predictor analyses.  

\subsection{Mortality}

We will report the number and proportion of participants who died during study follow-up, overall and by study arm. We will also describe the cause of death. Using the single-stage approach described above, we will test the null hypothesis that the SEARCH-Youth intervention did not reduce the two-year incidence of mortality with a one-sided test at the 5\% significance level. Additionally, we may visualize the survival curves with arm-specific Kaplan-Meier plots, pooling over clinics and censoring at outmigration, transfer, and database close.

\subsection{Documented transfers}

We will report the number and proportion of participants who formally transferred clinical care during study follow-up, overall and by study arm. Evidence of transfers will be documented from clinical records. Using the single-stage approach described above, we will test the null hypothesis that the SEARCH-Youth intervention did not increase the two-year incidence of transfers with a one-sided test at the 5\% significance level. Additionally, we may plot the estimated time-to-transfer in each arm, using the Kaplan-Meier method, pooling over clinics and censoring at death and database close.

\subsection{Care engagement}

We will conduct the following analyses, aiming to capture the multiple ways in which care engagement impacts participants outcomes. Overall and by study arm, we will report the number and proportion of participants who are 
\begin{packed_item}
\item \textbf{Engaged at two years:} having any contact with the clinic in the 6 months preceding the start of their endpoint window
\item \textbf{Retained throughout follow-up:} having contact with the clinic at least once every 120 days since enrollment
\end{packed_item}
where “clinic” refers to the original clinic where the participant enrolled in the study. 

For each of these endpoints, we will use the single-stage approach, described above, to obtain point estimates and statistical inference. We test the corresponding null hypotheses with a one-sided test at the 5\% significance level. In sensitivity analyses for engagement at two years, we will consider transfers as a success and censor at death. We will also provide clinic-specific estimates of care engagement and retention.

Pending data availability, we may also use the Kaplan-Meier method to estimate the longitudinal probabilities of \textbf{lapse}, defined as the first instance of a gap in contact for $>$120 days, with censoring at death, outmigration, transfer, and the start of the endpoint window. Using the two-stage approach, we will estimate the intervention effect on the probability of lapse by 6, 12, 18, and 24 months and formally test the null hypothesis.

\subsection{Severe adverse events}

Because the SEARCH-Youth trial does not include any drugs or clinical treatments, adverse events directly related to study intervention are not anticipated. However, we will track and report unexpected and significant adverse events. 

\section{Patient satisfaction surveys}

To assess mechanisms through which the intervention operates, we will administer a patient satisfaction survey during endpoint data collection. The survey aims to assess what works and what does not work for clients at their clinic.The survey covers potential barriers to and facilitators of care engagement and delivery. Specific dimensions include appointment convenience, ease of provider contact, alternative mechanism for clinic contact (i.e., phone visits), alternative mechanisms for drug delivery, trust in the provider, the provider’s ability to care for young people with HIV, and overall satisfaction with the clinic. Exact survey questions are available elsewhere. 

The survey questions will elicit Likert-type, evenly spaced responses, ranging from strongly disagree to strongly agree. In line with standard practice for such outcomes \cite{HayesMoulton2009}, we will evaluate them quantitatively. We will provide summaries of responses overall, by arm, and by clinic. We will formally test the null hypothesis that the SEARCH-Youth intervention did not reduce barriers to care delivery for adolescents and youth with HIV with a one-sided test at the 5\% significance level. To do so, we will use the two-stage approach, first calculating the clinic-specific, average response  and comparing these averages between randomized arms. Comparisons of these endpoints will be made on the absolute (i.e., difference) scale.   

\section{Intervention implementation}

Within the intervention arm only, we will also describe intervention implementation. Key metrics will include 
\begin{packed_item}
\item Number and proportion of participants who received 4+ life-stage assessments over the two-year follow-up period
\item Summaries of alternative choice access (e.g., off-site appointment, phone appointment, off-hours appointments, and off-site drug delivery), overall and by clinic
\item Median time to delivery of rapid viral load results, overall and by clinic
\item Proportion of viral load results delivered within 72 hours, overall and by clinic
\item Summaries of the counts and content of the provider e-collaborative chats
\end{packed_item}
All metrics will restrict to the primary study population: participants enrolled prior to December 1, 2019 and who remain in the region during the study period.

\section{Appendix} 

\subsection{History of changes}

This document does not cover additional analyses specified in the Study Protocol for cost-effectiveness or to evaluate the intervention effect on alcohol use or on HIV-free survival among infants born to participants. Separate statistical analysis plans will be developed for these endpoints.

\subsection{Power calculations}

Sample size and power calculations were based on the standard formulas for cluster randomized trials with proportion endpoints \citep{HayesMoulton2009}. We expect these calculations to be conservative, because of the precision gained through covariate adjustment in the analysis as well as our pre-specified choice of a one-sided hypothesis test.

We estimated 28 clinics (14 clinics/arm) would provide 80\% power to detect at least a 24\% relative increase in viral suppression at 2 years (the primary outcome) from 65\% in the standard-of-care, using a two-sided test and assuming a coefficient of variation of $k$=0.175 and at least 50 participants/clinic. As shown in the following Figure, these calculations are fairly insensitive to the number of participants. Under weaker clustering ($k$=0.125) or higher viral suppression in the standard-of-care arm, we anticipate being powered to detect smaller effects.

\begin{figure}[!h]
    \centering
\includegraphics[width=0.5\textwidth]{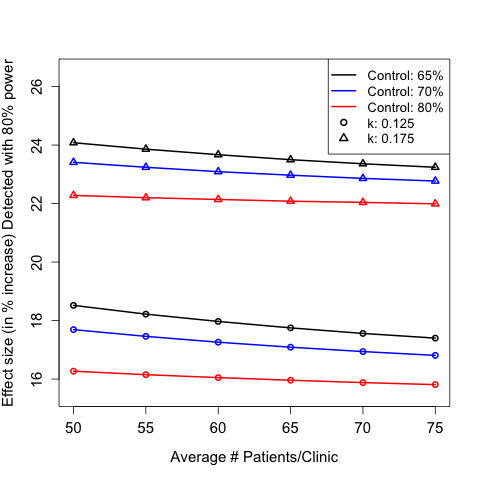}
\end{figure}

\bibliography{Bibliography}

\end{document}